\documentclass[structabstract]{aa} % este se utiliza para dos columnas
\usepackage{graphicx}
\usepackage{txfonts}
\usepackage{color} % Para colorear el texto
\begin{document}
   \title{Photometric distances to young stars in the inner galactic disk}

   \subtitle{II. The region towards the open cluster Trumpler 27 at L =
     355$^o$\thanks{Based on observations collected at Las Campanas
       Observatory, Chile},\thanks{Full photometric data and Tables 2 and 3 are available at the
       CDS via anomymous ftp to cdsarc.u-strasbg.fr (130.79.128.5) or
       via http://cdsarc.u-strasbg.fr/viz-bin/qcat?J/A+A/999/A999}}

   \author{Gabriel Perren\inst{1}
          \and
          Ruben A. V\'azquez\inst{2}
          \and
          Giovanni Carraro\inst{3}\fnmsep\inst{4}
          }

   \institute{Instituto de F\'isica de Rosario, IFIR (CONICET-UNR),
   Parque Urquiza, 2000 Rosario,       Argentina
              \email{perren@ifir-conicet.gov.ar}
         \and
                Facultad de Ciencias Astron\'omicas y Geof\'isicas (UNLP), Instituto de Astrof\'isica de La Plata (CONICET, UNLP), Paseo del Bosque s$/$n, La Plata, Argentina
               \email{rvazquez@fcaglp.unlp.edu.ar}
           \and
             ESO, Alonso de Cordova 3107, Casilla 19100 Santiago de Chile, Chile
             \email{gcarraro@eso.org}
             \and
             Dipartimento di Astronomia, Universita' di Padova, Vicolo Osservatorio 5, I$-$35122 Padova, Italy
             }

   \date{Received May 15, 2012 / accepted May16, 2012}

\abstract
  % context heading (optional)
{The spiral structure of the Milky Way inside the solar circle is
still poorly known because of the high density of the material
that causes strong extinction towards the galactic center.}
 % aims heading (mandatory)
{We present results of the first extensive and deep CCD photometric
survey carried out in the field of the open cluster Trumpler 27,
an object immersed in a region of extremely high visual absorption
in the constellation of Sagittarius not far from the Galaxy
center. The survey covers almost a quarter of square degree.}
 % aims heading (mandatory)
{We look for young stars clumps that might plausibly be associated
 with spiral structure. Wide-field $UBVI$ photometry combined with infrared information allows us to
reconstruct the distribution in reddening and distance of young
stars in the field using the Color-Color and Color-Magnitude diagrams}
%results heading (mandatory)
{The analysis of our data, combined too with extensive
spectroscopy taken from  literature shows that the real entity of
Trumpler 27 as an open cluster is far from being firmly stated. In
fact, instead of finding a relatively compact group of stars
confined to a small distance range, we found that stars
associated to Trumpler 27 are, indeed, a superposition of early
type stars seen along the line of sight extending over several
kiloparsecs beyond even the center of the galaxy. We demonstrate
that at each distance range it becomes possible to generate a
color-magnitude diagram resembling that of an open cluster. This
way, our analysis indicates that what was considered an open
cluster characterized by a significant age spread is a stellar
continuum that reaches its maximum number of stars at
approximately 3.5 kpc from the Sun, the distance of the
Scutum-Crux arm approximately. At the same time, and after
analyzing the way early type stars distribute with distance, we
found some of these stellar groups may be linked, within the
distance errors, with other inner spiral arms of our galaxy,
including the Near 3 kpc arm at approximately 5 kpc from
the Sun. However, very young stars by themselves do not seem to
trace strongly the inner spiral arms since they are distributed
evenly across several kiloparsecs toward the center of the Galaxy.
This is a remarkable difference with current HI and CO radio
observations maps that show inner spiral arms composed by discrete
structures of gas with a well defined inter-arm separation.} {}


  \maketitle

%
%________________________________________________________________

\section{Introduction}

One of the authors (Carraro 2011) has very recently examined the
parameters of the young diffuse stellar populations in the
direction $l = 314^\circ$ as seen in the background of some open
clusters and related them successfully with the inner
galactic structure. This study made use of photometric techniques
that we developed in the last decade and successfully applied to
the third quadrant of the Milky Way (Carraro et al. 2005; Moitinho
et al. 2006; V\'azquez et al 2008).\\

\noindent In Carraro (2011), hereafter the first paper of the
series, we extensively described the project and the employed
methods and proceeded to study the spatial distribution of early
type stars (spectral types O and B) in the direction of $l =
314^\circ$. Apart from the prominent Carina-Sagittarius spiral
arm, we detected for the first time signatures of the Scutum-Crux
arm  in that direction. These findings confirmed  the predictions
of earlier studies on the location and pitch angles of
part of the spiral arm pattern in the fourth Galactic quadrant.\\

\noindent The present article, the second of the series,
is aimed at improving our knowledge of the spiral structure in the
inner galactic disk in another direction, much closer to the
Galaxy center, at $l = 355^\circ$. In this case, instead of a
single pointing as in the first paper, we combined together three
pointings to cover a wide region ($\sim$ 25 arcmin on a side)  in
the direction of Trumpler 27. This is an open cluster located in
the Sagittarius constellation, 5 degrees from the center of the
Galaxy and slightly below the galactic plane, at coordinates
$\alpha_{2000}= 17h36m20s$, $\delta_{2000}= -33^\circ 31'$ ($l=
355^\circ.064, b= -0^\circ.742$). The zone occupied by the cluster
is heavily obscured by dust clouds but nevertheless it is still
possible to see a significant but sparse group of relatively
bright stars. The first extensive photometric survey on this
object was conducted by The and Stokes (1970) who obtained $UBV$
photographic photometry for about 40 stars spread over a circular
area of $11^{\prime}$ radius around the assumed center of this
cluster. They found huge values for the color excess $E_{(B-V)}$
ranging from -approximately- 0.6 to 2.5 mag. Despite this, they
found that the extinction law is normal with
$R_V=\frac{A_V}{E_{B-V}}=3.1$ and placed the cluster at a
heliocentric distance of 1.1 kpc. Later on, Moffat et al. (1977,
hereinafter MFJ77) conducted a deeper $UBV$ photometric survey
including photoelectric observations of the brightest stars and
photographic ones for the rest of them. This study was combined
with spectroscopic analysis for a significant number of stars
(about 50) and suggested that Trumpler 27 is at a distance of $2.1
\pm0.2$ kpc from the Sun with almost similar reddening values to
the ones previously computed by The and Stokes (1970). Bakker and
The (1983), used Walraven photometric data combined with infrared
and $UVB$ photometry. This last study confirmed that the
extinction law is normal, i.e. $R_V = 3.1-3.2$, that the
cluster is located at a distance of $1.7 \pm 0.25$ kpc and is about $10^7$ years old.\\

\noindent After a 20 years hiatus with no investigations, Trumpler
27 was the target of a new spectroscopic campaign carried out by
Massey et al (2001, hereinafter MDEW01) where the oldest spectral
types were verified and spectroscopy was made for several stars
that had not previously been measured. This new spectroscopic
survey revealed a surprisingly hot and massive stellar population
in the area of this object: one M supergiant, 10 supergiants of
B-types, two Wolf-Rayet stars, one Of-type star, two O-type giant
stars, other stars of B-types and G0I -probably- Cepheid
star. In MDEW01 the parameters of several stars
in Trumpler 27  were redetermined onto the basis of this new
spectral survey but retaining the old $UBV$ photometry made by
MFJ77. In this investigation the authors found that the cluster is
located at a distance larger than any previous determination: 2.5
kpc from the Sun. The next interesting finding made by MDEW01 is
the detection of a significant age spread amongst the B-type
stars. It appears that some of them show ages ranging from 6 to
$10\times10^{6}$ years being about 2 to
$4\times10^{6}$ older than the rest of them. They assume this age
spread comes from the large difficulty in separating cluster
members from background stars in this highly obscured zone.\\

\noindent Given its position -close to the Galaxy center
and onto the galactic plane- Trumpler 27 is an ideal target
for the purpose of investigating part of the spiral structure in
the fourth Galactic quadrant. Its surroundings contain a large
number of potential blue stars, some of which are considered
excellent tracers of spiral arms such as W-R stars and O-type
stars as shown by MDEW01 and MFJ77. It is useful to recall that it
has been almost 30 years since the last photometric study on this
object and that we are now in the position of taking advantage of
the large number of spectral types to re-discuss the distance and
age of Trumpler 27 with modern $UBVI$ observations. These two
parameters are of paramount importance because this cluster may
become a good galactic spiral arm tracer if the distance
discrepancy of the order of about 1.4 kpc from one author to other
is removed. In other words, such a difference means that the
cluster may belong to the Carina-Sagittarius arm or the
Scutum-Centaurus arm. But, first, we should assess whether
Trumpler 27 really exists and if the found distance
discrepancies are only due to the treatment of the strong
reddening of the region and/or to the fact that in
clusters so heavily obscured, like this one, only the
upper -vertical- main sequence is useful for fitting a reference
line (e.g., the ZAMS of Scmidt-Kaler (1982)).

In Section 2 we discuss the data acquisition, the reduction
process, the stellar astrometry and the comparison with former
photometry in the region. In Section 3 we show and discuss the
main photometric diagrams and introduce the strategy to derive
intrinsic parameters of stars in Trumpler 27. In Section 4, we
analyze the data set derived for hot stars and the nature of
Trumpler 27. Section 5 includes a description of findings
using infrared data. In Section 6 we collect the evidence to
interpret the structure of the inner Galaxy as seen through the
Trumpler 27 region. Conclusions are given in Section 7.

\section{Observation and data reduction}

Three regions containing Trumpler~27 were observed in the
$UBVI$ photometric system at Las Campanas Observatory (LCO) on
four different runs in 2006, 2010, and 2011, as illustrated in
Table~1, which lists useful details of the observations, like
filter coverage, airmass range and exposure time and sequences. We
used the SITe$\#$3 CCD detector onboard the Swope 1.0m
telescope\footnote{http://www.lco.cl/telescopes-information/henrietta-swope/}.
With a pixel scale of 0.435 arcsec/pixel, this CCD allowed us to
cover 14.8 $\times$ 22.8 arcmin on sky. In all runs the seeing was
good, ranging from 0.9 arcsec to 1.8. However, post-processing
indicated that in the 2006 and 2010 runs, the nights were not
completely photometric, and for this reason we only rely on the
2011 standard stars. The whole field of view is shown in Fig~1, a
montage of all V exposure images, and covers an area of 26
$\times$ 23 squared arcmin. North is up, and East to the left. To
determine the transformation from our instrumental system to the
standard Johnson-Kron-Cousins system, and to correct for
extinction, we observed stars in Landolt's areas PG~1323, PG 1633,
and MarkA (Landolt 1992) multiple times and with different
air-masses ranging from $\sim1.07$ to $\sim2.0$, and covering
quite a large color range -0.3 $\leq (B-V) \leq$ 1.7 mag. All the
images from previous runs have been then shifted to the 2011 run
by means of all common stars, before transforming the photometry
into the standard system.

\begin{table*}
\tabcolsep 0.5truecm \caption{$UBVI$ photometric observations of
Trumpler~27 and standard stars.}
\begin{tabular}{lcccccc}
\hline \noalign{\smallskip}
Target & RA & DEC & Date & Filter & Exposure (sec) & airmass (X)\\
\noalign{\smallskip} \hline \noalign{\smallskip}
Field-1 & 17:36:20.79 & -33:28:02.8 & 2006 May 31 & \textit{U} & 20, 200, 2x1500 &1.02$-$1.05\\
& & & & \textit{B} & 10, 100, 2x1200 &1.00$-$1.00\\
& & & & \textit{V} & 5, 10, 100, 900 &1.14$-$1.16\\
& & & & \textsl{I} & 5, 10, 2x100 &1.08$-$1.10\\
\hline

Field-1 & 17:36:20.79&-33:28:02.8 & 2006 Jun 26 & \textit{B} & 40 &1.00$-$1.01\\
& & & & \textit{V} & 20 &1.00$-$1.01\\
& & & & \textsl{I} & 20 &1.00$-$1.01\\
Field-1 & 17:36:20.79 & -33:28:02.8 & 2006 Jun 31 & \textit{U} & 60 &1.00$-$1.01\\
& & & & \textit{B} & 30 &1.00$-$1.01\\
& & & & \textit{V} & 30 &1.00$-$1.01\\
& & & & \textsl{I} & 30 &1.00$-$1.01\\
\hline
Field-1 & 17:36:20.79 & -33:28:02.8 & 2010 May 08 & \textit{U} & 15, 90, 300, 1500 &1.02$-$1.03\\
& & & & \textit{B} & 5, 45, 300, 1200 &1.06$-$1.10\\
& & & & \textit{V} & 2x5, 120, 900 &1.10$-$1.15\\
& & & & \textsl{I} & 2x5, 30, 600 &1.01$-$1.01\\
Field-2 & 17:36:54.79 & -33:28:02.8 & 2010 May 08 & \textit{U} & 15, 90, 300, 1500 &1.01$-$1.01\\
& & & & \textit{B} & 5, 45, 300, 1200 &1.03$-$1.04\\
& & & & \textit{V} & 2x5, 120, 900 &1.12$-$1.14\\
& & & & \textsl{I} & 2x5, 30, 600 &1.05$-$1.08\\
Field-3 & 17:35:47.79 & -33:28:02.8 & 2010 May 08 & \textit{U} & 15, 90, 300, 1500 &1.05$-$1.11\\
& & & & \textit{B} & 5, 45, 300, 1200 &1.11$-$1.18\\
& & & & \textit{V} & 2x5, 120, 900 &1.36$-$1.42\\
& & & & \textsl{I} & 2x5, 30, 600 &1.20$-$1.28\\
\hline
Field-1 & 17:36:20.79 & -33:28:02.8 & 2011 Jun 03 & \textit{U} & 60, 180 &1.00$-$1.01\\
& & & & \textit{B} & 30, 120 &1.00$-$1.01\\
& & & & \textit{V} & 10, 60 &1.00$-$1.01\\
& & & & \textsl{I} & 10, 60 &1.00$-$1.01\\
Mark A & 20:43:55.09 & -10:45:38.0 & 2011 Jun 03 & \textit{U} & 180, 2x240 &1.07$-$1.60\\
& & & & \textit{B} & 120, 2x180 &1.07$-$1.56\\
& & & & \textit{V} & 60, 2x90 &1.07$-$1.53\\
& & & & \textsl{I} & 60, 2x90 &1.06$-$1.51\\
PG 1323 & 13:25:38.91 & -08:50:05.6 & 2011 Jun 03 & \textit{U} & 3x180 &1.08$-$1.73\\
& & & & \textit{B} & 3x120 &1.08$-$1.77\\
& & & & \textit{V} & 3x60 &1.07$-$1.82\\
& & & & \textsl{I} & 3x60 &1.07$-$1.85\\
PG 1633 & 16:35:24.91 & 09:47:39.8 & 2011 Jun 03 & \textit{U} & 2x180, 240 &1.29$-$1.83\\
& & & & \textit{B} & 2x120, 180 &1.29$-$1.88\\
& & & & \textit{V} & 3x60, 90 &1.29$-$2.02\\
& & & & \textsl{I} & 2x60, 90 &1.29$-$1.97\\
\noalign{\smallskip} \hline
\end{tabular}
\end{table*}

\begin{figure}
\centering
\includegraphics[width=\columnwidth]{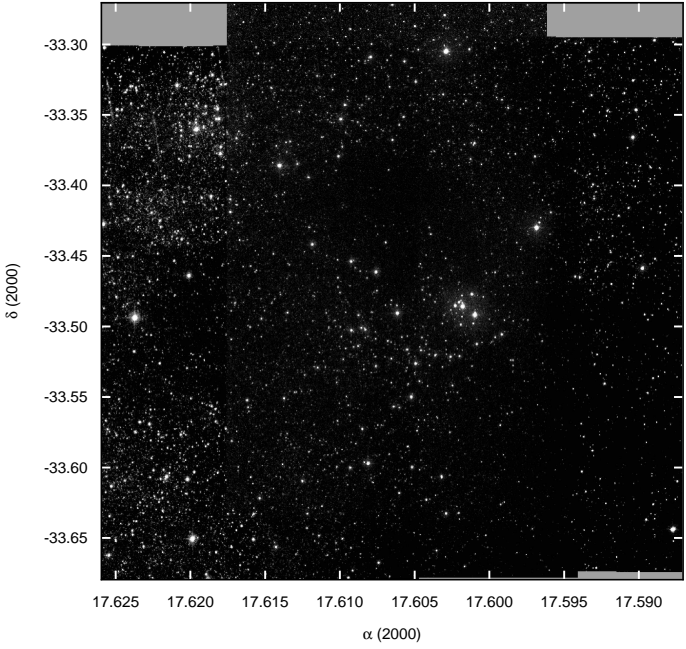}
\caption{Montage of all V images (28) taken in the area of
Trumpler~27. The field is about 26 arcmin $\times$ 23 arcmin.
$\alpha$ and $\delta$ coordinates are given in decimal notation.
North is up, East to the left. }
\end{figure}

\begin{figure}
\centering
\includegraphics[width=\columnwidth]{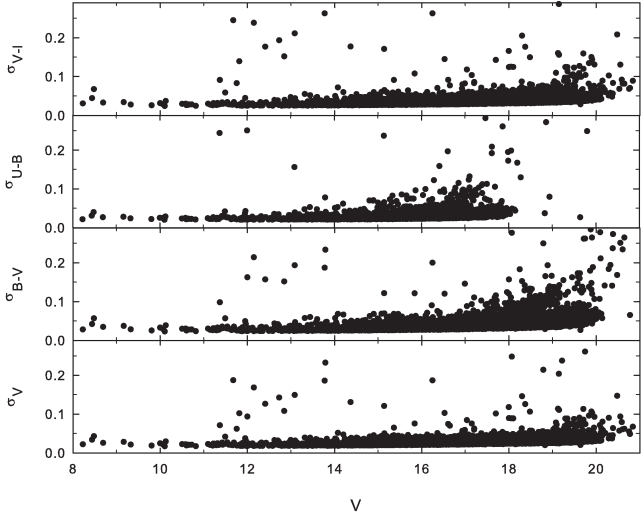}
\caption{Trend of global photometric errors in magnitude and
colors as a function of our $V$ magnitude. See text for details.}
\end{figure}

\begin{figure}
\centering
\includegraphics[width=\columnwidth]{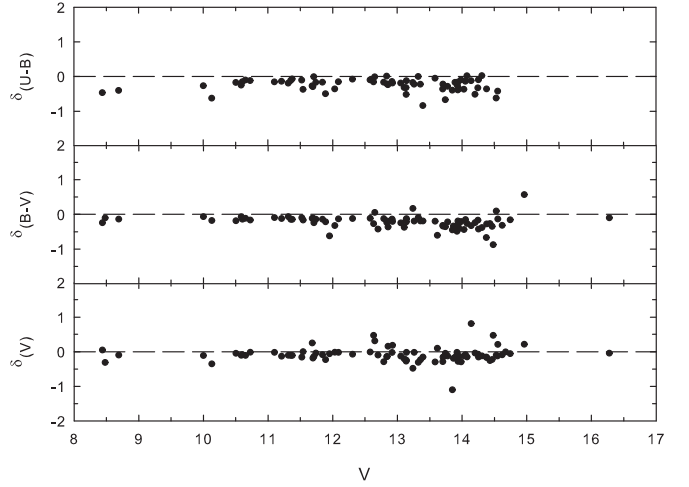}
\caption{Differences of our photometry with MFJ77 plotted against
our V magnitudes. Upper panel, $\delta_{U-B}$, mid panel
$\delta_{B-V}$, lower panel $\delta_V$.}
\end{figure}

\begin{figure}
\centering
\includegraphics[width=\columnwidth]{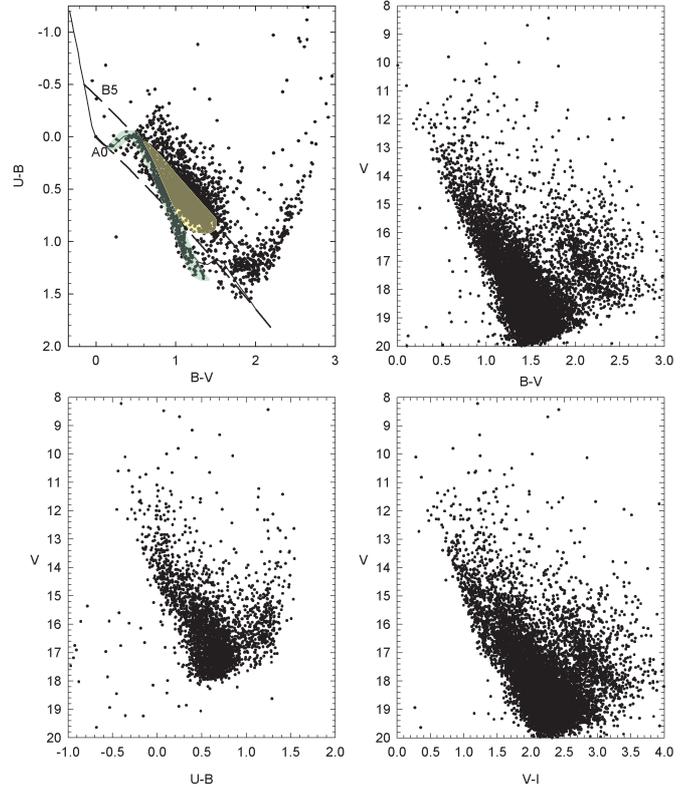}
\caption{The CCD -upper left panel- and the CMDs for $(B-V)$ -upper
right panel-, for $(U-B)$ lower left panel and for $(V-I)$ -lower
right panel of all observed stars. Solid line in the CCD is the
Schmidt-Kaler (1982) intrinsic line. Dashed lines in the
CCD mean the path of the reddening for a B5- and an A0-type stars
as indicated by the labels. The bluish region represents the
location of nearby dwarf stars; the yellowish one points out the
locus occupied by A-F-type stars affected by increasing reddening.
Stars above the B5 reddening line are young stars (see text),
except those ones for which no reddening solution is possible.}
\end{figure}

\subsection{Basic photometric reduction}
Basic calibration of the CCD frames was done using
IRAF\footnote{IRAF is distributed by the National Optical
Astronomy Observatory, which is operated by the Association of
Universities for Research in Astronomy, Inc., under cooperative
agreement with the National Science Foundation.} package CCDRED.
For this purpose, zero exposure frames and twilight sky flats were
taken every night. All frames were pre-reduced applying
trimming, bias and flat-field correction. Before flat-fielding,
all frames were corrected for linearity,
following the recipe discussed in Hamuy et al. (2006).\\
Photometry was then performed using the IRAF DAOPHOT/ALLSTAR and
PHOTCAL packages. Instrumental magnitudes were extracted following
the point-spread function (PSF) method (Stetson 1987). A
quadratic, spatially variable, master PSF (PENNY function) was
adopted, because of the large field of view of the detector.
Aperture corrections were then determined making aperture
photometry of a suitable number (typically 15 to 20) of bright,
isolated, stars in the field. These corrections were found to vary
from 0.160 to 0.290 mag, depending on the filter. The PSF
photometry was finally aperture corrected, filter by filter.

\subsection{Photometric calibration}
After removing problematic stars, and stars having only a few
observations in Landolt's (1992) catalog, our photometric solution
for the run was extracted combining measures from both nights-
after checking that they were stable and similar- yielding
a grand total of 63 measurements per filter, and turned out to be:\\
\noindent
$ U = u + (5.004\pm0.010) + (0.49\pm0.010) \times X + (0.129\pm0.013) \times (U-B)$ \\
$ B = b + (3.283\pm0.006) + (0.25\pm0.010) \times X + (0.040\pm0.008) \times (B-V)$ \\
$ V = v + (3.204\pm0.005) + (0.16\pm0.010) \times X - (0.066\pm0.008) \times (B-V)$ \\
$ I = i + (3.508\pm0.005) + (0.08\pm0.010) \times X + (0.037\pm0.006) \times (V-I)$ \\
\noindent
where $X$ indicates the airmass.\\
The final {\it r.m.s} of the fitting in this case was 0.040,
0.019, 0.015, and 0.015
in $U$, $B$, $V$ and $I$, respectively.\\
\noindent Global photometric errors were estimated using the
scheme developed by Patat \& Carraro (2001, Appendix A1), which
takes into account the errors resulting from the PSF fitting
procedure (i.e., from ALLSTAR), and the calibration errors
(corresponding to the zero point, color terms, and extinction
errors). In Fig.~2 we present our global photometric errors in
$V$, $(B-V)$, $(U-B)$, and $(V-I)$ plotted as a function of $V$
magnitude. Quick inspection shows that stars brighter than $V
\approx 20$ mag have errors lower than $\sim0.05$~mag in magnitude
and lower than $\sim0.10$~mag in $(B-V)$ and $(V-I)$. Higher
errors, as expected, are seen in $(U-B)$. The final catalogue
contains 9769 \textit{UBVI} entries.

\subsection{Astrometry}
The optical catalogue was cross-correlated with 2MASS, which
resulted in a final catalog including \textit{UBVI} and
\textit{JHK$_{s}$} magnitudes. As a by-product, pixel (i.e.,
detector) coordinates were converted to RA and DEC for J2000.0
equinox, thus providing 2MASS-based astrometry, useful for {\it
e.g.} spectroscopic follow-up.The rms of the residuals in the
positions were 0.$^{\prime\prime}$15, which is about the
astrometric
precision of the 2MASS catalogue (0.$^{\prime\prime}$12, Skrutskie et al. 2006 ).\\

\subsection{Comparison with former photometry}

Up to now, the most extensive catalogue of $UBV$ photometry in the
region of Trumpler 27 belongs to MFJ77. The other photometric sets
are only photographic or in the Walraven system. Comparing our
data set with MFJ77 we found 83 common stars in $V$, 81 in $(B-V)$
and 68 in $(U-B)$. The differences in the sense of our data minus
MFJ77, are shown in Fig. 3 as a function of our $V$. Most of the
stars in the MFJ77 survey have photographic photometry and just a
few have photoelectric photometry. In the comparison process, we
found at least half a dozen stars showing strong differences in
apparent magnitude and/or color indices. This may be produced,
specially in crowded regions, in an unavoidable star-light
contamination by nearby neighbors when using diaphragms in
photoelectric measures. Mean differences and standard
deviations are $\overline{\delta_V} =-0.08\pm0.15$, $\overline{\delta_{(B-V)}} =
-0.19\pm0.15$ and $\overline{\delta_{(U-B)}} = -0.23\pm0.17$. The systematic
trend in Fig. 3, in the sense that our values are brighter
and bluer than those of MFJ77 and the large scatter, is likely
produced by some zero point differences between our photometry
and MFJ77's photographic and photoelectric photometry made with
different telescopes and different diaphragm apertures,
respectively.

\section{The all area photometric diagrams}

Earlier CCDs and CMDs of the region of Trumpler 27 have been
always very difficult to interpret because  of
strong and variable internal reddening and the consequent difficulties in
defining the upper main sequence. Besides, the upper part of
Trumpler 27 CMD is quite vertical and short what makes any fitting
of the ZAMS (Schmidt-Kaler 1982) dubious. The sum of
these difficulties have precluded this object from a more in-depth
study so far (McSwain \& Giess, 2005).

Figure 4 shows the CCD (upper left panel), the $V$ vs $(B-V)$ (upper
right panel), the $V$ vs $(U-B)$ (lower left panel) and the $V$ vs
$(V-I)$ (lower right panel) CMDs. The solid line in the CCD is the
ZAMS of Schmidt-Kaler (1982) shown in the normal $R_V=3.1$ position.
The reddening path for a B5-type
star is also indicated. A simple visual inspection of the CCD allows to establish
three dominant stellar populations:

\begin{description}

\item $\bullet$ a first one, above the
reddening line for a typical B5-type star, composed undoubtedly
of blue stars strongly affected by variable reddening;

\item  $\bullet$  a second stellar population appears below the reference line for a
B5 star (the yellowish zone) composing a sort of stellar
band starting at the second knee of the ZAMS and extending to the
red side because of different amounts of reddening affecting them.
This large star concentration corresponds, photometrically
speaking, to A and early F-type stars but also to reddened
B-type stars later than B5 as indicated by the path of the
reddening in the CCD;

\item  $\bullet$  finally, a third stellar population (greenish coloured) little affected by
reddening appears extending from the domain of A to M stars all
along the intrinsic reference line. Most of these stars
are likely to be nearby objects since visual absorption maps from
Neckel and Klare (1980) reveal a sudden absorption increment at
less than 1 kpc in this direction. Therefore, the detection of
intrinsically faint stars not reddened or just marginally affected
by interstellar dust is only possible if they are close enough to
the Sun.

\end{description}

\begin{figure*}
\centering
\includegraphics[width=\textwidth]{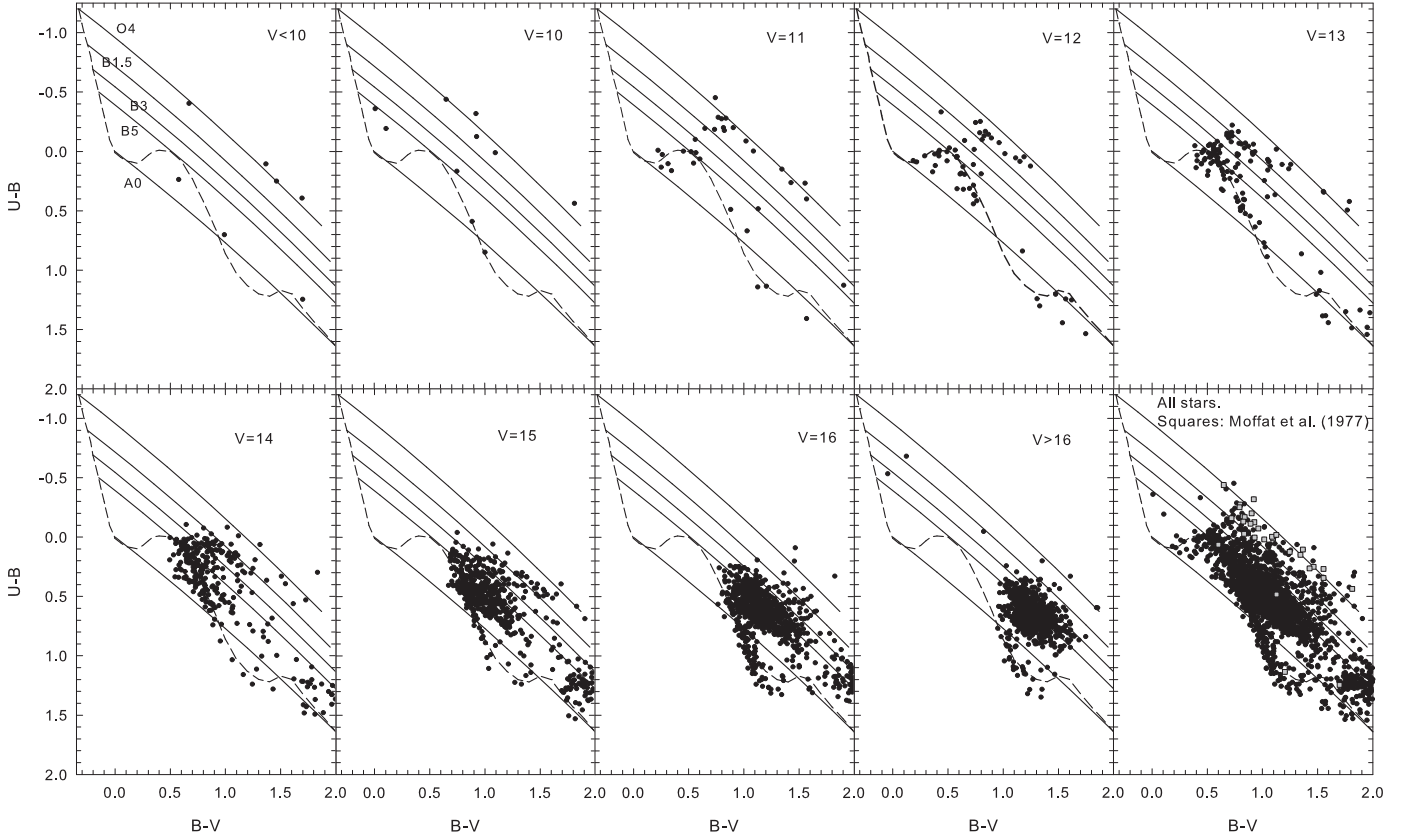}
\caption{The CCDs of the segmented CMD at intervals of 1 mag.
Dashed line shows the position of the Schmidt-Kaler (1982)
intrinsic line. The succession of solid lines show the path of the
reddening for some of the early type stars indicated with the
respective labels. The panel in the right lower corner shows the
overall stars and those with spectral types, grey squares, from
MFJ77 }
\end{figure*}

Inspection of  Fig. 1  lends further support to this scenario:
intense absorption is seen at the west-southwest and immediately
north of the center of Trumpler 27,  and also to the east. A clear
star overdensity is seen, in the same figure, at the northeast and
at the expected location of the cluster core.

The CMDs [$V$ vs $(B-V)$, $V$ vs $(U-B)$ and $V$ vs $(V-I)$] are less
illustrative since not much can be extracted from them beyond the
presence, above V = 16 mag, of an extended upper stellar sequence
with a high degree of color dispersion that reflects the strong
influence of dust in the zone. Since the three populations
discussed previously are undistinguishable from each other in the
CMDs, we estimate this type of diagram plays a marginal
role to inferring useful information on this region.

\begin{table*}
\tabcolsep 0.25truecm \caption{Intrinsic photometric values,
distances and spectral types of stars in Trumpler 27}
\begin{tabular}{lcccrcclrl}
\hline \noalign{\smallskip}
Id  &   $E_{(B-V)}$   &   $(B-V)_0$   &   $(U-B)_0$   &   $M_V$ &   $d$   &   $\sigma_d$  &   ST$^1$  &  Id(MFJ77)  &   ST$^2$(MFJ77,MDEW01)  \\
\noalign{\smallskip} \hline \noalign{\smallskip}
5517    &   1.69    &   -0.28   &   -1.00   &   -2.60   &   3.17  &   0.73  &   B0.5    & (...) & (...)    \\
4732    &   1.10    &   -0.28   &   -1.00   &   -2.60   &   2.64  &   0.58  &   B0.5    &   19  &   OB-     \\
5355    &   1.14    &   -0.21   &   -1.02   &   -6.90   &   6.60  &   0.29  &   B0.5Ia  &   16  &   B0.5Ia  \\
6456    &   1.12    &   -0.02   &   -0.62   &   -7.10   &   6.74  &   0.15  &   B0Ia    &   2   &   B0Ia    \\
5672    &   1.80    &   -0.27   &   -0.95   &   -2.35   &   3.38  &   0.78  &   B1      & (...) &  (...)    \\
\noalign{\smallskip} \hline\noalign{\smallskip} \hline
\end{tabular}
\newline
\begin{minipage}[]{16cm}
Id first column is our star numbering. Id column nine is the MFJ77
numbering. ST$^1$ indicates the photometric spectral type found in
this paper. ST$^2$(MFJ77,MDEW01) indicates spectral types from the
literature.
\end{minipage}
\end{table*}

\begin{table*}
\tabcolsep 0.25truecm \caption{Idem for stars outside Trumpler 27}
\begin{tabular}{lcccrcclrl}
\hline \noalign{\smallskip}
Id  &   $E_{(B-V)}$   &   $(B-V)_0$   &   $(U-B)_0$   &   $M_V$ &   $d$   &   $\sigma_d$  &   ST$^1$  &  Id(MFJ77)  &   ST$^2$(MFJ77,MDEW01)  \\
\noalign{\smallskip} \hline \noalign{\smallskip}
442     &   1.22    &   -0.25   &   -0.90   &   -2.10   &   3.41  &   0.84  &   B1.5    & (...)   & (...)  \\
4497    &   1.09    &   -0.25   &   -0.90    &   -2.80    &   2.83   &   1.33  &   B1.5:V: &   32  &   B1.5:V: \\
3629    &   1.15    &   -0.26   &   -0.95   &   -3.20    &   3.22   &   0.95  &   B1:V:   &   34  &   B1:V:   \\
8932    &   1.25    &   -0.26   &   -1.00  &   -5.40    &   2.64   &   0.40  &   B1II    &   103 &   B1II    \\
\noalign{\smallskip} \hline \noalign{\smallskip} \hline
\end{tabular}
\newline
\begin{minipage}[]{16cm}
Columns as in Table 2.
\end{minipage}
\end{table*}

\begin{figure*}
\centering
\includegraphics[width=\textwidth]{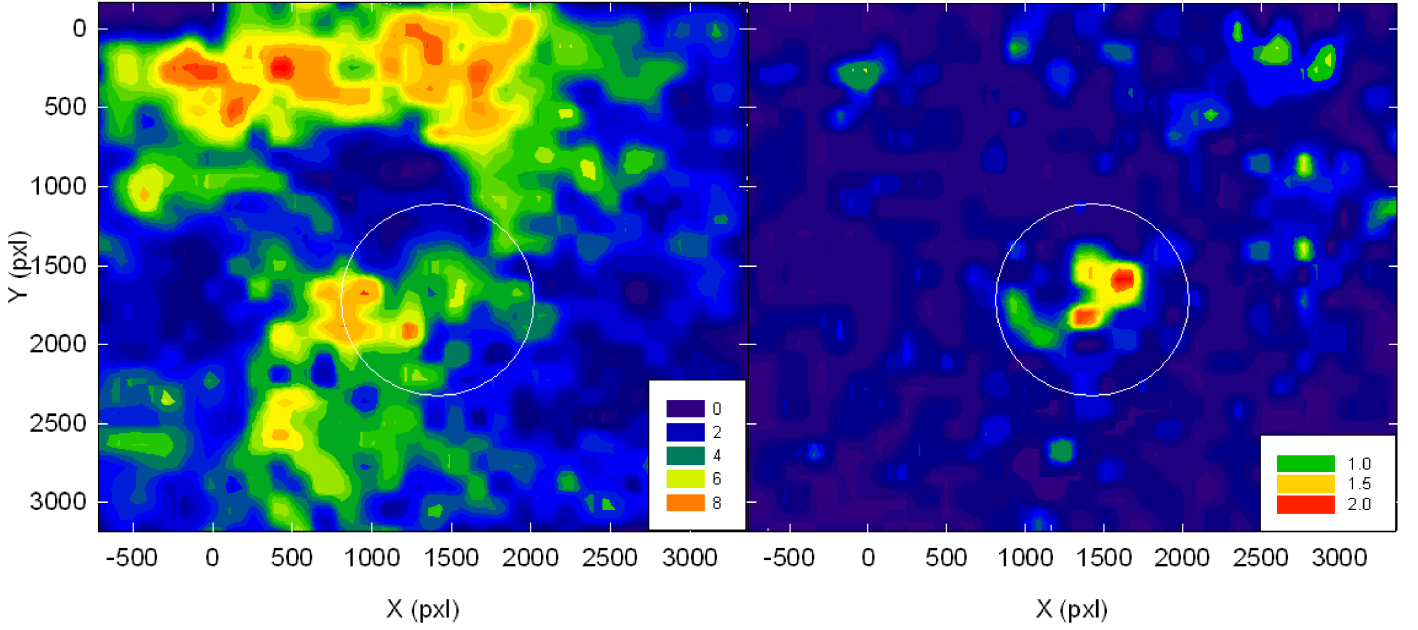}
\caption{The left panel is the contour plot showing the
star density in the surveyed area generated with 9000 stars
approximately. The insert gives the number of stars per
half-minute squared box. The right panel shows the contour plot of
those stars early than B5-type (see text). The
insert as in the left panel. The 600 pxl circle indicates the
region assumed to be the core of the cluster.}
\end{figure*}

\begin{figure*}
\centering
\includegraphics[width=\textwidth]{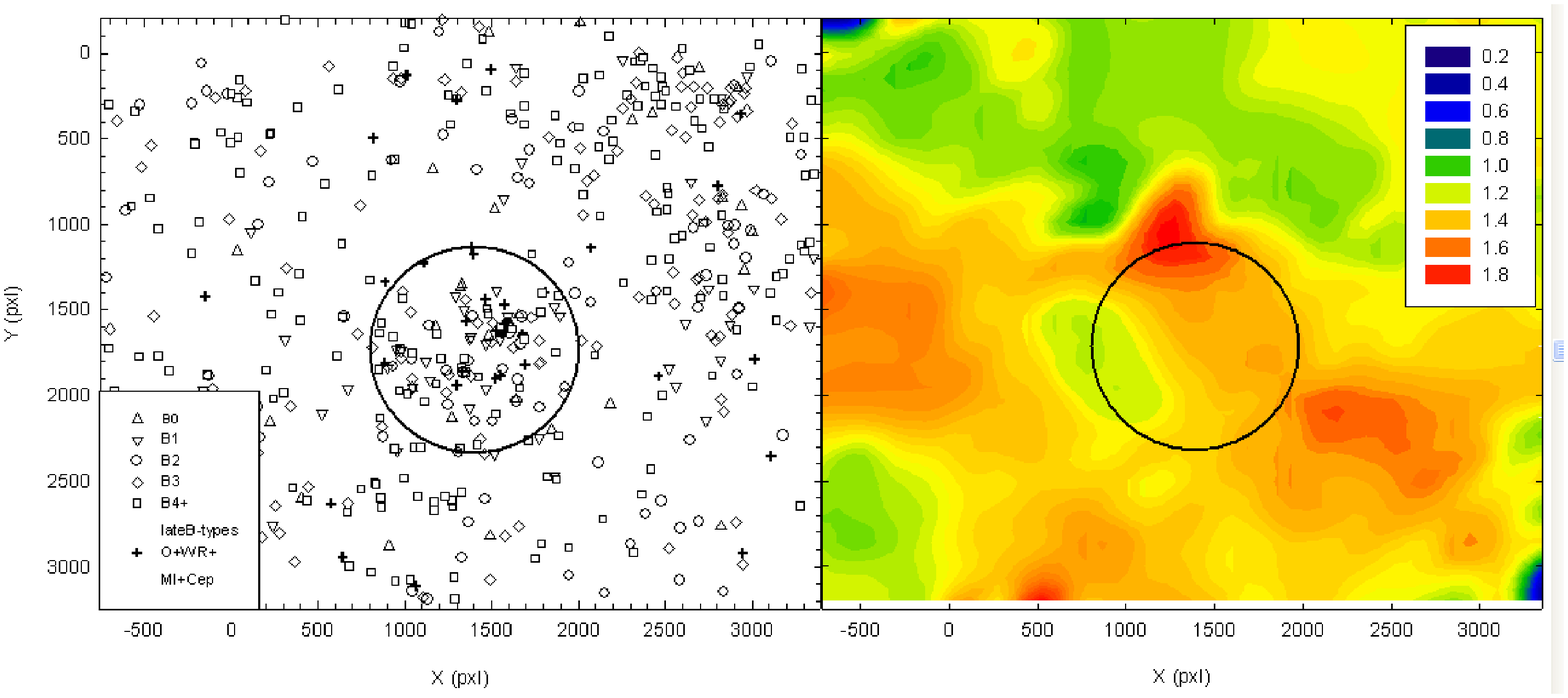}
\caption{The left panel stands for the spatial distribution of
early type stars according to their spectral types -shown in the
insert- as explained in the text. The right panel is the contour
plot of $E_{(B-V)}$ across the cluster surface. The insert gives the
amount of $E_{(B-V)}$. The circle has the same meaning as in Fig. 6.}
\end{figure*}

\subsection{Deriving the parameters of early-type stars}

If the CMD is segmented with magnitude intervals and the
corresponding CCD to each interval is built, it becomes simpler to
understand the way stellar populations develop in a given zone.
The foundation and applications of this kind of analysis
in open clusters and Galactic fields is developed in Carraro et
al. (2008, 2009). In the present case the procedure is shown in
the series of CCD in Fig. 5 at intervals of 1 mag down to $V =
18$. The superimposed lines follow the slope $ E_ {(U-B)}
= 0.72 \times E_ {(B-V)} + 0.05 \times E_{(B-V)}^2 $ explained
long time ago by Hiltner and Johnson (1956) (corresponding to
$R_V=3.1$) for stars of various spectral types from O to B5 in
addition to the A-type stars.

The series of panels shows the color range covered by the three
above mentioned star populations. Actually, reddened hot stars
along the reddening lines for O and B4 types look well detached
from the rest down to $V\sim14$. A comparison of our CCD with the
ones produced by Bakker and The (1983), in their Fig. 3, or in
Fig. 5 of MFJ77, suggests that the major difference, apart from
the inclusion of many faint stars, rests in the high number of new
hot and reddened stars detected in our field which in turn is
larger than the others.

The panel in the right lower corner shows the overall CCD where we
have identified, with open squares, bright stars already
studied in previous articles via spectral analysis.

From $V\sim14$ a stellar group (yellowish coloured) composed of a
wide strip of A and F-type stars becomes evident reaching the
maximum at $V=15$. Certainly, this latter group shows an ample
range of reddening. Beyond this limit dwarf stars of G-K- and even
some M-type stars are detected at increasing number all
along the intrinsic line indicating they are not so much
affected by reddening and belong to the greenish zone. As
stated above, these stars and part of the less reddened A-F-
stars in the group belong to a nearby population in the line of
sight to Trumpler 27.

In the remaining of this paper we shall focus on those
stars above the reddening line for B5-type stars as indicated in
Fig. 5 to derive individual distances by means of the
spectroscopic parallax method since we know they have unambiguous
reddening solution in the CCD. This method has proved to be
very efficient to approximate spectral types from
photometry alone when unique reddening solutions are available as
shown in Ahumada et al. (2011). Intrinsic colors for each
star in this zone of the CCD were derived using the reddening
line slope $E_{(U-B)} = 0.72\times E_{(B-V)}+0.05\times
E_{(B-V)}^2$ along with the relation $(U-B)_0 = 3.69
\times(B-V)_0+0.02$ (Carraro et al. 2010). This gives a second
order equation where the $(B-V)_0$ index is the positive root and
the $(U-B)_0$ index comes from the above expression. This locates
the star on a point in the ZAMS (Schmidt-Kaler 1982) where the
nearest spectral type is assigned. Unlike the Q-method
that derives the $(B-V)_0$ star color assuming a linear relation
of color excesses, here we adopt a quadratic relation that
provides a more accurate solution when the color excess is high.
Once individual color excesses and intrinsic colors are derived,
we proceed to assign spectral types and absolute magnitudes
following the relationship given by Schmidt-Kaler (1982). In this
procedure all stars are assumed to be of luminosity class V.
However, a good number of hot stars have spectral types and
luminosity class from MDEW01 and MFJ77. In these cases, the
assignment of intrinsic color and absolute magnitude depend on the
spectral type and luminosity class found by these authors using
again the Schmidt-Kaler (1982) calibrations. Individual
distance moduli are given after correcting apparent magnitudes by
$A_V= 3.1\times E_{(B-V)}$. We accept $R=3.1$, which is
completely in line with the rigorous study of the absorption
throughout Walraven and infrared photometry developed by Bakker
and The (1983) in the region.

The distances obtained are photometric and are subject,
therefore, to uncertainties depending primarily on photometric
errors. As done in Carraro (2011), we analyze the impact of these
errors on distances by starting with the well known
expression for the distance modulus:

\begin{equation}
V-M_V = -5 + 5 \times \log (d) + A_V
\end{equation}

\noindent A simple error propagation leads to:

\begin{equation}
 \epsilon_{ (d)} = \ln(10) \times d \times 0.2 \times [ \sigma_V + 3.1 \times
\sigma_{(B-V)} ]
\end{equation}

\noindent Assuming that $\epsilon_{(A_V )} = 3.1 \times
\epsilon_{(B-V)}$, where $\sigma_{(V)}$ and
$\sigma_{(B-V)}$ directly come from photometry. In the final
error computation we adopt $\epsilon_{ (M_V )} =0$.

The results of this procedure yielded about 600 stars with
intrinsic parameters and distances as indicated in Tables 2 and 3
in a self explanatory format. Inspecting the errors in distance in
the same tables, it follows that typical errors are about 20\%,
quite reasonable for the method and of the same order of the
obtained by the usual ZAMS (e.g. Schmidt-Kaler 1982)
superposition. We want to emphasize that the assignment of
spectral types entirely rests on the star positions onto the CCD
(Carraro 2011) as if they were single stars. We recall that some
of them may be undetected binaries or unresolved double stars what
introduces another source of distance uncertainty due to altered
colors and magnitudes. Even if we deal with true single stars
having MK spectral types this is not enough to reduce
uncertainties in distances since the MK classification and
therefore the assigned $M_V$ may vary from author to author (see
cases of different classification in Tables 2 and 3). Finally, it
is evident that even amongst the stars located along the reddening
line for B5-type stars there is still a chance of contamination by
intrinsic data scatter of some reddened A-type stars. Therefore,
and in an attempt to minimize wrong star inclusions, we
restrict the spectroscopic parallax method to those stars in the
range from B4.5 to O-types (unless otherwise stated). We
are confident that, despite the errors, our large data sample will
clearly reveal the main properties of the stellar populations in
the region of Trumpler 27.

\subsection{Discussion on spectral types and distances of some stars}

Here we  briefly discuss the nature and parameters of a few
particular stars listed in Tables 2 and 3 for which distances have
been derived using the spectroscopic parallax method. We focus
mainly on some stars for which MK spectral types and
thin-prism classification comes from the literature. We also
discuss briefly some potential variable stars. We refer to
all these stars
using the identification given in MFJ77.\\

{\bf Star 1}: MDEW01 confirm the oldest spectral type, M0Ia, given
by MFJ77. It is also a suspected variable (NSV 22849) as indicated
by Samus et al. (2010). The distance of 1.8 kpc has been
derived under the assumption that it is not a variable. Additional information
on this objects is commented later when using infrared data.\\

{\bf Star 2 (LSS 4253)}: the old spectral type from MFJ77, O9Ia,
turns out to be B0Ia according to MDEW01 with no indication about
membership. This last spectral type and our new
photometry locate it at almost 7 kpc from the Sun, outside any
reasonable distance limit
given for Trumpler 27 in the literature.\\

\noindent {\bf Stars 8, 10, 12, 19, 20, 21, 25 and 30}, all have
thin-prism plates classification in MFJ77 (a
classification technique that indicates whether a star is a
potential OB star) and none of them were studied spectroscopically
by MDEW01. In most of these cases they were classified as
potential OB stars with different degrees of accuracy as seen in
Tables 2 and 3. Our procedure confirms the MFJ77 thin-prism plates
classification approximation as they resulted of photometric
spectral types O4, O7, O8.5, B0.5, B1.5, B1, O4 and O6-types, respectively. \\

\noindent {\bf Star 16 (LSS 4263)}: the early spectroscopy gave
O9.5 II: but changed to B0.5 Ia after MDEW01. The new spectral
type and our photometry locate it at a distance of about 7 kpc.\\

\noindent {\bf Star 23}: The spectral type determined by
MFJ77, B0.5Ia, was changed to B0.7Ia after MDEW01. This star is
also known as V1082 (Samus et al. 2010), a probable eclipsing
variable. We derived a distance of 3.5 kpc but if it becomes a
true binary star, the distance could be reduced to 2.6 kpc in the
most favorable case of similar masses.\\

\noindent {\bf Star 44}: this star is a B1.5Ia according
to MDEW01 who have given concluding spectroscopic
arguments favoring it is a supergiant star and not a giant
one as stated in MFJ77. In view of the distance of 2.5 kpc they
computed for Trumpler 27, they assumed this star is not a cluster
member, an assertion we agree with because the distance we obtain
with our photometry is about 16 kpc from the Sun. Even if
we assume it is a binary star its distance would
still be more than 10 kpc away from the Sun.\\

\noindent {\bf The WR stars}: on one side, star 28 (WR 95) was
classified WN5 by MFJ77 and WC9 by MDEW01; we adopted here this
latest and modern classification. On the other side,  star 105
(HDE 318016) is a WC7/WN6 following MDEW01. Conti and
Vacca (1990) reported distances of 2.8 kpc for the WC9 and 2.4 kpc
for the WC7/WN6 that were adopted by MDEW01 in their analysis. To
deal with these two complex stars we proceeded to derive their
intrinsic $(B-V)$ colors by means of the relationship between the
Smith (1968) $ubv$ and $UBV$ systems as given in Lundstr\"{o}m \&
Stenholm (1984). According to the relationship given in this last
paper, it was assumed $(b-v) = -0.4$ for the WC9 (WR 95) and
$(b-v) = -0.3$ for the WC7/WN6 (105) while the absolute
magnitudes were taken from the newest values given by van der
Hucht (2001). Excesses of color obtained using the relationships
of Smith (1968) were $E_{(B-V)} = 1.95$ for WR 95 and 1.4 for star
105. Finally we derived distances of 2.2 for WR95 and 3.5 kpc for
star 105 that we will use in the analysis next section.\\

\noindent {\bf Star 107 (LSS 4257)}: it was classified B0V by
MFJ77 but the new spectral type from MDEW01 drastically changes
the point by setting this star as a supergiant of a later type,
B0.5Ia. As in the previous case of star 44, they concluded
this is another non-member star. We agree since our distance
sets this star at more than 10 kpc from the Sun. Even in case of
binarity, its distance would be reduced to 7.6 kpc,
too far from any previous estimate of the cluster distance.\\

\noindent {\bf Star 102 (V925 Sco, HD159378)}: this is a
variable star with no well established period from $70^\texttt{d}$
to more than $300^\texttt{d}$ and light variation of a few tenths
of a magnitude (Samus et al. 2010). Since, according to van
Genderen (1980), it may not be a Cepheid variable of Pop I, we
treated it as a G0 supergiant star and derived a distance of 2.4
kpc not far from the 2.1 kpc given by MFJ77. Parthasarathy \&
Reddy (1993) placed this object at 3 kpc from the Sun but they
gave no precise indication on the way they got their reported
distance.

\section{Hunting for Trumpler 27}

We made an attempt to use proper motions in the area of
Trumpler 27 from the UCAC3 data-base (Zacharias et al. 2010). The
putative distance (at 2-3 kpc from the Sun) and the crowding of
the region towards the Galactic center made the exercise useless.
Errors in both proper motion components are large and no feature
can be detected. This is exactly as in Moni Bidin et al. (2011),
where three possibly new globular clusters are discussed toward
the Galactic bulge using data from the VISTA VVV survey (Minniti
et al. 2011). As in that case, proper motions are so scattered
that we even restrain from showing the results.

In previous papers we have explained (see e.g. V\'azquez et al.
2010) that instead of using star density profiles to set the
cluster boundaries we prefer to look for the zone of higher star
density in a contour plot. Indeed, density profiles reduce
an open cluster -irregular in shape by definition- to a centrally
peaked spherical stellar density distribution. On the contrary
contour plots reveal details of the impact of absorption on the
cluster shape and allow to define the boundaries in a more
realistic approach. The procedure works well in areas where the
absorption is high like the one in this paper. As in the case of a
radial density profile we set the size of a cluster to the area
enclosed when the contour plot reaches a flat value.
Notwithstanding and for the sake of completeness of the present
analysis, in the next section we shall examine the infrared
density profile. Meanwhile, Fig. 6 (left panel) shows the contour
plot ($0.5\times0.5$ arcmin box using the Kriging gridding method)
generated with more than 9000 stars detected in the region of
Trumpler 27. The full contour plot shows no star overdensity at
all in the place where the cluster is assumed to be (the bright
stars near the center of the mosaic in Fig. 1).
Overdensities are detected north and south of the location
of the cluster center confirming what a simple eye-inspection
shows in Fig. 1. Actually, the lack of a notorious overdensity
right there suggests that Trumpler 27 has been detected for being
handful of bright stars well detached from the rest by dust
clouds. This is better shown in the right panel of Fig. 6 which
displays the contour map generated only with all early-type stars
of Trumpler 27 included in Tables 2 and 3 (spectral types down to
B4.5). The circle in this panel has 600 pxl radio, just over 4
arcmins and surrounds entirely the weak overdensity emerging at the
position of the cluster. That is, the contour map in the right
panel shows that the cluster is just a handful of early stars
separated by a thick dust ring from other similar groups in the
region. Figure 7, left panel, displays stars in Tables 2 and
3 according to their spectral types, describes clearly the above
assertion. Note that the stellar density in this region cannot be
assumed different from that north-west and west. The role played by
the absorption is also well depicted by the contour map of
$E_{(B-V)}$ excesses built in the right panel of Fig. 7 with the
values listed in the tables. Here we see that reddening varies in
the range from 1 to more than 2 mag across the surface with higher
values in a wide band from the south- south-west to the north-east
that separates the diverse groups seen in the left panel. A
comparison of the right panel in Fig. 7 and the left panel (all
stars) in Fig. 6 suggests that a vast fraction of the A-type stars
surveyed in the area may be immersed in the dust and just a few of
them together with others of later types are likely located in
front of the dust clouds. Otherwise, the overdensities seen north
and south-east in the left upper panel should disappear in this
plot.

\begin{figure}
\centering
\includegraphics[width=\columnwidth]{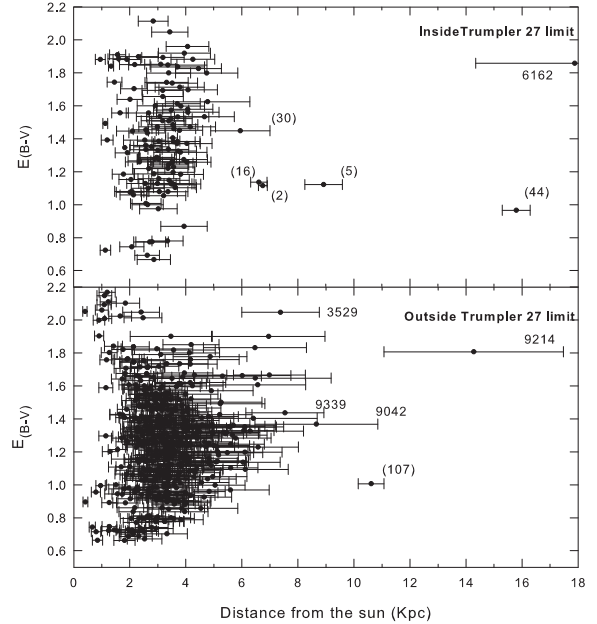}
\caption{The path of reddening with distance for stars inside
(upper panel) and outside (lower panel) 600 pxl radius.
Horizontal bars are distance errors computed from (2).
Some of the distant stars are indicated. Numbers in parentheses
mean the MFJ77 notation. }
\end{figure}

The zone of Trumpler 27 was included by Bitran et al.
(1997) in a CO survey focused in a dozen degrees around the Galaxy
center. The cluster is not only immersed in a region of large CO
emission but, most interesting, it is projected against a zone of
extremely high absorption as shown in map 222 made by Neckel and
Klare (1980). This alone justifies the appearance of our CCDs in
Fig 5 where potential A-F-type stars start with little absorption
to end up, at $V= 16$, completely immersed in the dust clouds.
Likewise, the latest spectral type stars in the field have to be
necessarily nearby stars since they are intrinsically too
faint and do not appear much affected by reddening.

Figure 8 traces the path of reddening with distance inside
and outside the cluster boundaries. Error bars coming from Eq. (2) are
also shown. It is worth mentioning briefly the impact of distance
errors into our analysis. We discussed in previous section the
meaning of distance errors which are mainly based on the
propagation of photometric errors which are random in nature.
Fig. 8 shows the trend of our sample of stars
toward the Galaxy center and we do not see that errors can modify
any of our conclusions. When looking through the core of the
Trumpler 27 region the bulk of early type stars extends for more
than 4 kpc while some others may reach 9 kpc and even more than 15
kpc as indicated. Something similar happens outside the 600 pixels
limit with some stars located at large distances too. The
reddening behavior shown in this figure is well in line with the
above mentioned absorption map 222 from Neckel and Klare (1980).
Map 222 and nearby ones show that absorption increases after 0.5
kpc from the Sun, rises up to 5-6 mag in the next kiloparsec and
keeps high up to 4-5 kpcs towards the Galaxy center. Our deeper
and extended sample of young stars not only confirms findings of
Neckel and Klare (1980) but also states strong absorption
variability all across the region with $0.65 < E_{(B-V)} < 2.2$
mag meaning more than 6 mag of visual absorption. Fig. 8 shows too
that dust clouds are not a local phenomenon but they are presents
for several kiloparsecs starting near the Sun and spanning over 4
kpc (at least) towards the center of the Galaxy.

Histograms for different spectral types as a function of distance
for both regions are shown in Fig. 9. Left and right panels are
for stars inside and outside Trumpler 27, respectively. The upper
panels include O-. WR-, B0- and late type supergiants. Mid panels
is for B1-B2- type stars and lower panels is for B3-B4- types. For
the sake of an easy visualization, we restricted the plot to all
the stars up to 7 kpc from the Sun.

In terms of confirmed spectral types, we found that 19 stars out
of 126 inside the cluster limits have some type of spectral
classification (MK or objective prism) from Massey and/or MFJ77
including O- and B-type stars and one WC9 star; the supergiant
star No 1 belongs to this zone. Outside the cluster radius there
are 501 early type stars; 9 of them, including the G0 Cepheid
variable, have spectral analysis: one WC7/WN6, one O8.5III and one
early B-type supergiant.

Inside the cluster limits O-type stars are seen nearly at all
distances. The rest of stars show a narrower distribution beyond
1.5 kpc from the Sun. Late dwarf B-types are lacking in the first
1-1.5 kpc probably because of dust but also because of an
statistical artifact. The lack of them at larger than 4-5 kpc is
mainly due to a combination of extinction and intrinsic faintness
of these spectral types. In the left panels there seems to
be two peaks, one at $\sim$1-1.5 kpc and the other -higher- at
$\sim$3.5 kpc. The 1-1.5 kpc peak is not far from the estimated
distances by The and Stokes (1970) and Bakker \& The (1983) for
Trumpler 27 although it is a bit far, however, from the 2.1 and
2.5 kpc given by MFJ77 and MDEW01, respectively. A more
refined analysis indicates this peak contains about 20 stars
including three O-type stars at $\sim 1.5$ kpc. and six other hot stars
with distances between 0.9 and 1.6 kpc. The rest of the stars of
photometric B-types (eight of them have evolved MK types) spread
over distances from 1 to 2 kpc. Three of them with confirmed
evolved MK types are in a distance range between 1.1 and 1.7 kpc.
The M0Ia supergiant star is placed at 1.8 kpc in our distance
estimation but since colors of extremely red stars may be quite
uncertain it is likely located at the same distance than the others.
The WC9 star, placed at 2.2 kpc, is unlike to be related with this
star group. The mean distance of all the stars in this stellar
grouping (allowing for distance errors) is $1.7\pm0.4$ kpc.
This is almost 1 kpc below the value derived by MDEW01. Compared
to other distances computed for Trumpler 27, it is 0.5 kpc less
than the one estimated by MFJ77, 2.1 kpc, and about 0.5 kpc larger
than another value given by The \& Stokes (1970), 1.1 kpc, but is
in good agreement with the distance published by Bakker \& The
(1983), 1.6 kpc from the Sun. However, the simple grouping by
itself -with huge distance spread- does not indicate the presence
of a true open cluster as we shall see later on. If we look
outside the region of the location of Trumpler 27 in Table 3 we
find only 35 stars sharing a similar distance range.

\begin{figure}
\centering
\includegraphics[width=\columnwidth]{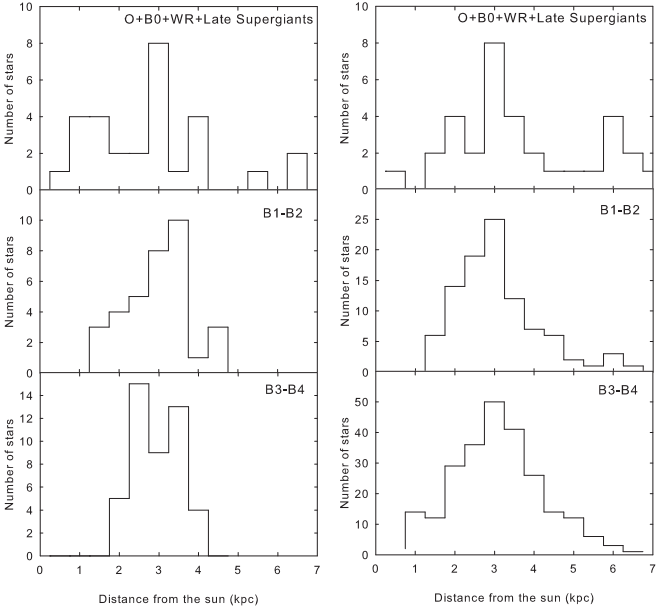}
\caption{Histograms of stars for spectral type. Left panels for
those inside the cluster limits and right panels for those stars
outside the cluster (see text for explanation).}
\end{figure}

Regarding the second star peak at $\sim$3.5 kpc seen in
the direction to Trumpler 27 core, it is quite far from
the distance 2.1 and 2.5 kpc given for Trumpler 27 by MFJ77 and
MDEW01. Moreover, it is wider than the first one and tends to show
some kind of coincidence with the evident peak of stars outside
the cluster region at the same distance approximately (see right
panels in Fig. 9). In this distance range from 3 to 3.5 kpc there
are more than 110 stars outside the potential cluster core
including 3 stars of O-types and one B-type supergiant out of 27
stars with MK types. The WC7/N6 and the G0I stars, both at 2.4 kpc
approximately, are probably not related with this second peak of
stars either.

\begin{figure}
\centering
\includegraphics[width=\columnwidth]{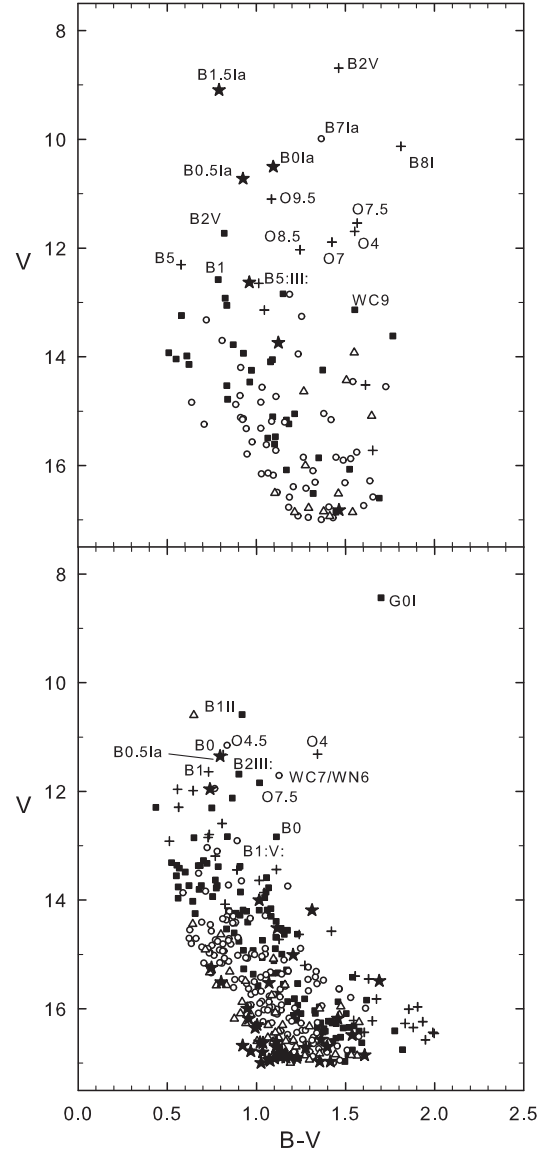}
\caption{CMDs of stars ordered by distance. Symbols are: (+) stars
from 0 to 2 kpc, ($\blacksquare$) for 2-3 kpc, ($\circ$) for 3-4
kpc, ($\vartriangle$) for 4-5 kpc and ($\bigstar$) stars beyond 5
kpc. Upper panel, all stars above B5-spectral types in Trumpler 27
region. Lower panel, idem above for stars outside.}
\end{figure}

The histograms of Fig. 9 are conclusive in the sense that early
type stars are found at any distance inside and outside the
cluster region. That is why, in our opinion, different authors
arrive at different conclusion about members and distances of this
putative open cluster. An additional proof is provided by the
observed CMDs of early type stars found inside and outside the
cluster boundaries shown in Fig. 10. Here stars have been
identified with different symbols according to their distances and
immediately arises that any combination of them may produce a
believable CMD. In particular, the stars in the 3-4 kpc range are
responsible for the left envelope and high density of both CMDs
while the most dispersed stars are those belonging to the 2-3 kpc
range. To be remarked is that inside the limits, where it is
assumed Trumpler 27 to exist, the upper part of the CMD is
composed by stars as near as 2 kpc and as far as more than 5 kpc
from the Sun.

From the above analysis we conclude then that the evidence  is
not conclusive enough to clearly establish the true nature and
location of Trumpler 27. There is no doubt here that we are dealing
with a diffuse young population of stars distributed all along the
line of sight with some stellar clumps including all types of
bright stars.

\section{Infrared information}

Infrared data provided by the GLIMPSEII spring'08 (highly
reliable) catalogue was cross-correlated with our optical
catalogue to get 2MASS $JHKs$ and IRAC 3.6, 4.5 and 8 $\mu$m data.
In particular, $JHKs$ data was used to perform an additional search for
Trumpler 27 as can be seen in Fig. 11 (down to $J= 16$ mag). Using all stars in
the GLIMPSEII catalogue which lie inside the boundaries of our observed frames, a radial density profile was constructed as shown
in the upper left panel of Fig. 11. The profile
gives the number of stars per square arcmin computed as the number
of stars found in concentric annuli 1 arcmin wide starting at the
current adopted cluster coordinates given in \S 1. Error bars are
Poisson errors (computed as $\sqrt N$ with $N$ the number of
stars per square arcmin). The panel confirms that the cluster is
only an artifact produced by the evident dust ring around the
center as shown by the pronounced dip at 3 arcmin from the center.
This is the same shown in Fig. 6, left panel. Apart from the dip
the star density profile is essentially plain, within the Poisson
errors, showing no evidence of a star clustering.

Close to star No 1 and 1b (MDEW01 numbering) there exists the
maser source OH355.1-0.76 measured by Knapp et al. (1989) in their
circumstellar CO emission survey. Unfortunately they give no
further information on this object since and it was discarded due
to the strong contamination caused by galactic CO emission over a
wide range of velocity. We assume that, probably, the supergiant
star No 1 is the optical counterpart of this object since that, in
turn, is also a suspected variable (Samus et al. 2010). Across the
area surveyed in this article we found optical counterparts of two
YSO candidates (Nos 5338 and 5269 in our numbering, corresponding
to 2MASSJ17362441-3331289 and 2MASSJ17362515-3325521 sources,
respectively) and another star (No 1796, our numbering) with huge
infrared excess.

\begin{figure}
\centering
\includegraphics[width=\columnwidth]{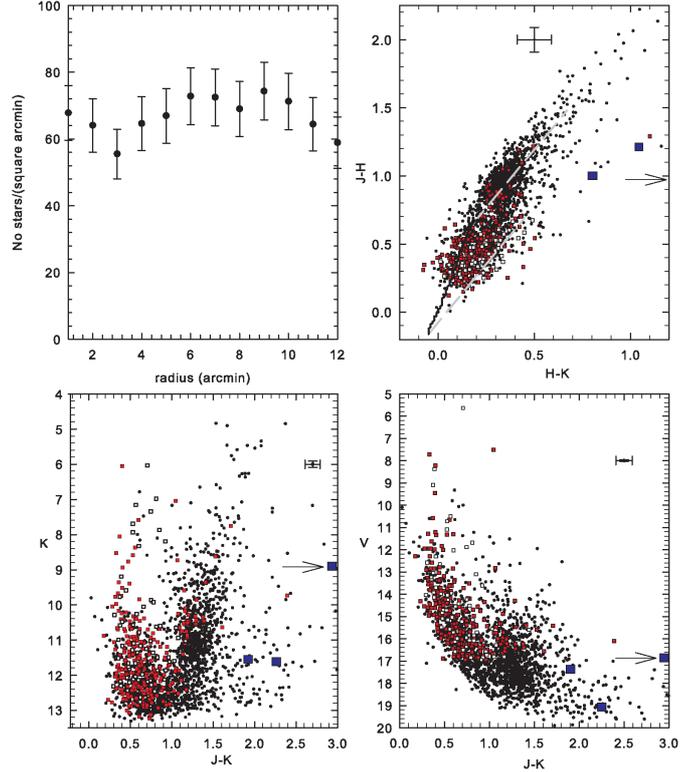}
\caption{The radial density profile in the region of Trumpler 27 -upper left panel. Bars are Poisson
errors. The $(J-H), (H-K)$ color-color diagram of all stars in the
region is shown in the upper right panel. White and red squares
are stars in Tables 2 and 3 respectively. The cross indicates the
average color errors. Large blue squares give the position of the
YSO candidates and the arrow points out the position of the star
No 1796 outside the diagram limits. Solid black line is the
intrinsic relation from Koornneef (1983); the two grey dashed
lines are color excesses relations for $A_V= 6$ mag for early and
late type stars. The lower left and right panels show the $K, (J-K)$ and $V, (J-K)$ color-magnitude diagrams. Symbols as in the
upper right panel.}
\end{figure}

Upper right panel in Fig 11 is the overall $(J-H), (H-K)$ two color
diagram. The intrinsic relation from Koornneef (1983), solid black
line, and the 6 mag reddening vectors $E_{(J-H)}/E_{(H-K)} = 1.7$, gray
dashed lines, for early and late type stars are indicated. We have
superimposed stars earlier than B5 inside 600 pxls radius (white
squares) and outside these limits (red squares). Despite the fact that data
scatter is strong and typical infrared errors are large, this
diagram confirms that stars in the region of Trumpler 27 are
affected by strong variable reddening and shows no indication of a
clear cluster sequence. Indeed, the spread of the data inside 600
pxl is similar to the one shown by the other early stars outside
this limit. Large blue squares and the arrow show the position of
the YSO candidates above mentioned and the star No 1796. It seems
that more objects in the zone exhibit large infrared excesses. The
other two lower panels are the $K, (J-K)$ and $V, (J-K)$
color-magnitude diagrams. These two diagrams show the same pattern
found in Fig. 4 in the sense that no clear cluster sequence
emerges from them and no evidence of a pre-main sequence is
revealed. The illusory vertical strips seen in the $K, (J-K)$ are
composed by stars at very different distances inside the 600 pxl
radius and also outside that limit. Actually, with no other
information available to impose boundary conditions (as the one
provided by the CCD in Fig. 4 and MK types) it would be possible
to confuse the vertical strip with the upper sequence of an unreal
cluster. For the sake of completeness we have also plotted 3.6,
4.5 and 8 $\mu$ data in the adequate diagrams (not shown to save space) and, again, no
evidence for a cluster has been found in any of them.

\section{The region toward $l = 355^\circ$ and the inner spiral structure of the Galaxy}

The presence of over 600 stars of very early spectral
types in a small sky area of a quarter of square degree located in
a continuous range of distance up to 7 kpc at least from the Sun
is an exciting result of our survey that connects this
young diffuse population to  the inner Galactic structure. In the
traditional model of four arms for the galaxy (Russeil 2003,
Vall\'ee 2005) moving from the Sun location inwards -in
the region of Trumpler 27- there should be Sagittarius-Carina arm,
Scutum-Crux arm, Norma and the Near 3 kpc-arms at 0.6, 3, 4 and 5
kpc respectively. In the proposed recent view face-on of our
galaxy (Churchwell et al. 2009), the cross with Sagittarius-Carina
arm should happen at 1.6 kpc. In particular, this arm looks weak
and narrow; this is, the arm is tiny and not very prominent. In
the same sketch the  Scutum-Crux arm is located at 3.2 kpc and
becomes the most prominent feature of the inner galactic
structure. The Norma arm is another tiny structure intersected at
4.4 kpc and, finally, the Near 3 kpc arm at over 5 kpc is the less
pronounced of all the features. A recent map of the Milky Way has
been presented by L\'epine et al. (2010) where the traditional
four arm structure is abandoned and replaced by polygonal arms
described by molecular CS associated with IRAS sources. Given the
complex structure proposed by L\'epine et al. (2010) we prefer to
compare our findings with the traditional description (e.g.
Russeil 2003).

In terms of the stellar components, mainly O-B3-type stars and
open clusters associated with the grand design spiral structure,
the information is scarce as shown by Russeil (2003). The OB type
stars catalogue recently compiled by Reed (2003) has been used by
Carraro (2011) to highlight the spiral structure around the Sun.
Fig.1 in Carraro (2011) demonstrates that only O-type stars show
some sort of clear coincidence with spiral arms of the Galaxy
while B-type stars, on the contrary, show a smooth distribution up
to the reach of the survey made by Reed (2003). Carraro
(2011) also emphasizes the lack of coincidence of the star sample
(O- plus B-types) with the trace of Scutum-Crux. In the direction
we are dealing with, the Reed (2003) catalogue scarcely reaches 5
kpc from the Sun while the number of confirmed (spectroscopically)
O-type stars by Ma\'iz-Apell\'aniz et al. (2004) -complete only to
$V=8$ mag- indicates that very few confirmed stars of this type
are found in the direction where Trumpler 27 is located.

Returning to our data sample we have included stars of Tables 2
and 3 in the distance vs $E_{(B-V)}$ plot of Fig. 12 up to 18 kpc
from the Sun. In the figure we also have marked the
distances to which the spiral arms of the Galaxy should be located
approximately according to the current accepted model of four arms
and their respective widths. We briefly recall that while arms
separation is of the order of 3 kpc the width of them is quite
uncertain. We set arms widths in Fig. 12 using the value of about
1 kpc given by Vall\'ee (2005). Accepting that Sagittarius-Carina
is at less than 1 kpc in the $355^\circ$ direction we found very
few stars associated with this spiral arm, mostly O and
B3-B4, while the bulk of young stars tend to gather at the
distance where Scutum-Crux and Norma arms are set. The plots in
Fig. 12 are in line with Fig. 1 shown by Carraro (2011) in the
sense that very young stars do not define conclusively the trace
of inner spiral arms but  are distributed evenly across
several kiloparsecs toward the center of the galaxy. This is a
singular difference with radio observations. In fact, maps of the
HI and CO contributions suggest that these arms are discrete and
well outlined structures, Scutum-Crux being one of the major of
them moving toward the inner galaxy.

\begin{figure*}
\centering
\includegraphics[width=\textwidth]{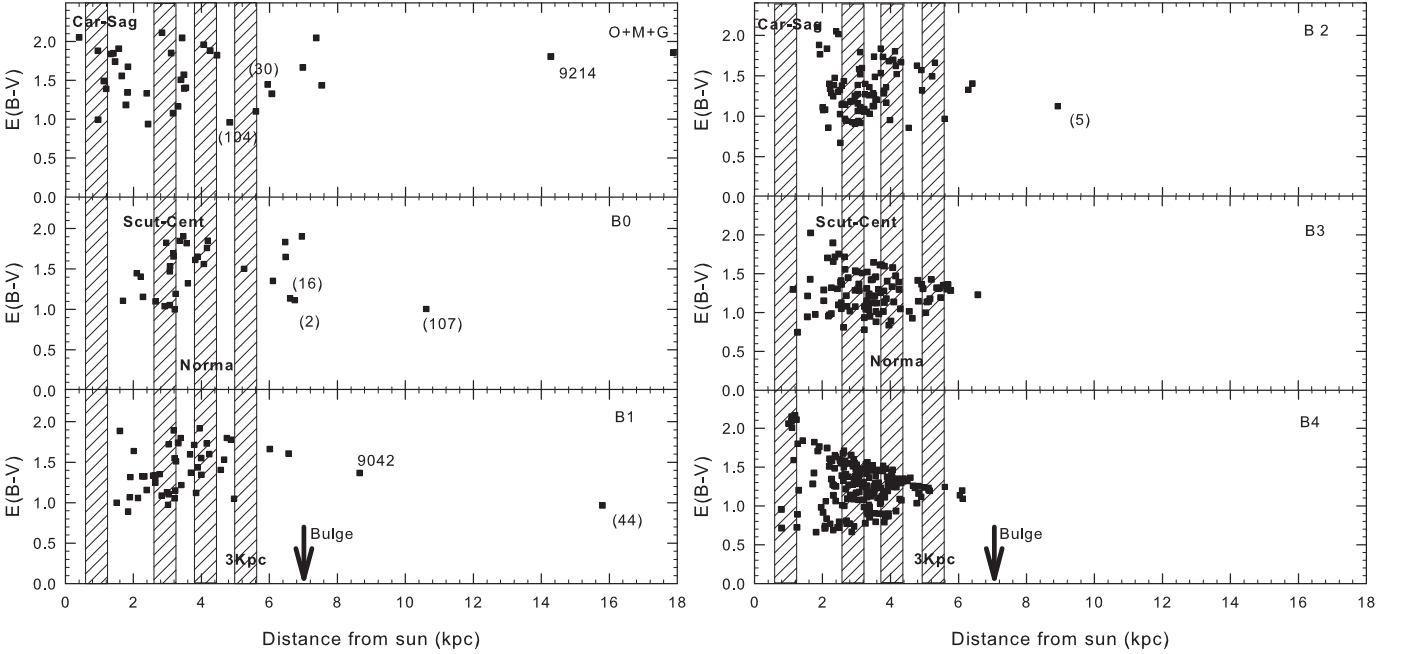}
\caption{The path of reddening with distance according to the star
spectral type. The approximate position and width of the inner
spiral arms is shown by dashed columns; the probable beginning of
the Bulge is indicated by the arrow. Numbers in parentheses as in
Fig. 8.}
\end{figure*}

The recent study by Dame \& Thaddeus (2008) collected specific and
irrefutable evidence of the presence of the distant counterpart of
the Near 3 kpc arm based in CO and previous observations in HI.
The authors estimated that distances from the Sun for both arms
are 5.1 and 11.8 kpc for the Near 3kpc arm and the Far 3 kpc arm,
respectively. It is worth mentioning that stars
at all spectral types (also one O-type) can be found in or close
to the Near 3 kpc arm in Fig. 12. If the tendency in our sample is
correct, for the first time we are looking at components of
a very recent star formation process there. From the point of
view of our findings the relation between the young population and
the Near 3 kpc arm is not controversial. As Dame \& Thaddeus
(2008) show, the Near 3 kpc arm extends mostly below the
galactic plane at the location of Trumpler 27 and therefore, part
of the young stars in our sample may be spatially connected with
this arm. There are stars even farther in this region (at more
than 8 kpc) that have no chance to be related with the Far 3 kpc
arm since it extends mainly above the galactic plane.

From our study, based on an extended sample of early type stars
detected photometrically, we find it hard to establish the
relation of the inner spiral arms with our stellar population.
While the gas component is expected to trace the path of
every inner arm by means of discrete structures of molecular CO
and HI, stellar distances of young stars do not. Fig. 12 allows us
to assure that the trend of the young diffuse population extends
all along 7 kpc towards the center of the Galaxy but by no means
can we specify the position and distance from the Sun of each arm.
This difficulty arises from a combination of distance
uncertainties with geometrical circumstances. In fact, despite
the process of star formation is active and vigorous it is hard to
relate it to specific patterns of spiral structure because of
a) the tight packing of four arms in such a short
distance range and b) our line of sight on the plane is
almost at right angles with the spiral structure.

All the circumstances mentioned above are reflected in the radio
observation domain. Very recently, Green et al. (2011) have been
studying the distribution of methanol masers in the inner galaxy
covering the longitude range from -28$^\circ$ to +28$^\circ$ in
longitude. In particular, in the direction of our analysis, they
found an enhancement of methanol sources which are comprehensible
only as a superposition of contributions from large galactic
structures confused by intervening spiral arms along the line of
sight. This difficulty is exactly the same we found in this paper.

\section{Conclusions}

This is the first time that an $UBVI$ CCD photometric study is
carried out in the region of the open cluster Trumpler 27 towards
$l=355^\circ$. In order to determine the properties of the stellar
population, we have applied well known broad-band photometric
techniques. Our goal has been twofold: on one  side to recognize
the main parameters of the cluster -if it exists- since
several observers in the past arrived at different conclusions
and, on the other side, to analyze the field star population
present in the region, since the same studies underlined the
presence of a large number of bright stars that have no relation
with the cluster itself.\\

\noindent

An open cluster is, in essence, a decreasing sequence of
gravitationally bound stellar masses well confined in a volume of
space. In the case of nearby open clusters, the analysis of proper motions and radial velocities becomes a powerful tool to asses
the cluster membership on any star found in the field.  As
for distant clusters, the way to assess if a star belongs to it is
by resting on a spectrophotometric study focused on some sort of
stellar overdensity detected against the sky background. In the
zone of Trumpler 27 there is neither reliable proper motions
-given the distances of the stars in the region- nor appreciable
star overdensity even using infrared data as shown in Fig. 11,
upper left panel.

From a spectrophotometric point of view we cannot confirm
or rule out definitely the existence of the cluster since results of our analysis based on separating early type
stars from field stars are not conclusive at all. This is, CMDs
showing the same pattern can be constructed with stars at
different distances from the Sun neighborhood all the way to the
Galactic center. Massive stars of O-type, evolved B-types, M- and
G- super-giants plus two WR stars are found at all distances and
widely separated -angularly- from each other making it unlikely to
connect them, from an evolutionary point of view. The different
CMDs in optical and infrared are typical of very young star groups
superposed along the line of sight resembling, e.g., OB
associations. This could be the reason for the age spread detected
by MDEW01 since the entire star group associated historically with
Trumpler 27 is composed by stars which did not form in a same
event and do not share the same spatial location.

Basing on our data, it is difficult to assess the existence of
Trumpler 27 at the distance, with the members and with the age
derived in previous articles already cited. In addition, we were
unable to find in our diagrams a star sequence composed by faint stars
that could be related beyond doubt with potential massive members of Trumpler 27.
This is a serious obstacle and we find it unlikely that photometry alone could solve
the question in the future. Certainly, the controversy may be
settled by means of an exhaustive spectroscopic study able to reach
also faint stars. An interesting point to interpret, in terms of
star formation too, is the lack of any evident HII region in the
zone as expected in a region of massive star formation.
WR and O-Type stars are usually injecting a tremendous amount of
ultraviolet photons in the surrounding gas which should be seen in
emission but there are no clear traces of emitting gas.\\

\noindent
As far as the diffuse young star population is concerned,
in the direction $l=355^\circ$, our star sample shows an even
distribution.
This suggests that early type stars define  a continuous superposition
of stars  with no density peaks from one arm to the next, except for
the prominent clump at the location of the Scutum-Crux arm,
as seen in Fig. 9.

The findings of this paper can be connected with the ones from
Carraro (2011) at $l=314^\circ$. Firm evidence was found there of a young stellar
population associated with Carina-Sagittarius and Scutum-Crux
arms. Moreover, the existence of an even star distribution in the
direction $l=314^\circ$, where arms are more open than in the Galactic
center direction, was found. This confirmed the outcome of old
(e.g. Muzzio and Levato 1980 and references therein) and more
recent (Carraro \& Costa 2009, Baume et al. 2009) studies that
provided robust optical evidences of their existence and their
potential as spiral arm indicators.

However, in the present study which looks towards the center of
the Galaxy, four spiral arms are closely packed in  only 5 kpc
from the Sun. Therefore, the task of disentangling one structure
from the next is more cumbersome, and would require the support of
an improved radio trigonometric parallax method to measure
accurate distances and proper motions of obscured star forming
regions in the Milky Way (Brunthaler et al. 2011).

Meanwhile, we believe it is of extreme value to continue our
program which aims at studying the inner disk stellar populations
using $UBV$ and infrared photometry by extending it to
other absorption windows, as in the case of Trumpler 27. Young
stellar populations can be very well tracked using $UBV$
photometry and when combined with findings from radio
observations, a better picture of large structures can be achieved
certainly.

\begin{acknowledgements}
G. Carraro expresses his gratitude to the Las Campanas Observatory
staff, and in particular to Patricio Pinto for the excellent
support during various observing runs. We also thank Sandy Strunk
for reading carefully the manuscript. G. Perren and R.A. V\'azquez
acknowledge the financial support from the CONICET PIP1359. This
research has made use of the NASA/IPAC Infrared Science Archive,
which is operated by the Jet Propulsion Laboratory, California
Institute of Technology, under contract with the National
Aeronautics and Space Administration.

We acknowledge useful and constructive comments from an anonymous referee
which allowed us to improve the manuscript.
\end{acknowledgements}

\end{document}